\documentclass[12pt]{article}

\usepackage{psfig}
\newcommand{\be}{\begin{equation}}
\newcommand{\ee}{\end{equation}}
\newcommand{\ba}{\begin{eqnarray}}
\newcommand{\ea}{\end{eqnarray}}
\newcommand{\BE}{\begin{equation}\label}
\newcommand{\BEQ}{\begin{eqnarray}\label}
\newcommand{\EE}{\end{equation}}
\newcommand{\EEQ}{\end{eqnarray}}
\newcommand \beq{\begin{eqnarray}}
\newcommand \eeq{\end{eqnarray}}

\newcommand{\DK}[1]{\mbox{\boldmath$#1$}}

\newcommand{\bs}{\begin{small}}
\newcommand{\es}{\end{small}}

\newcommand{\bi}{\begin{itemize}}
\newcommand{\ei}{\end{itemize}}
\newcommand{\om} {\omega}

\begin{document}

\title{Dynamic structure factor and collective excitations of neutral and Coulomb fluids}

\author{J.Ortner}

\date{to be published in Phys. Scripta}
\maketitle
\begin{center}
{Institut f\"ur Physik, Humboldt Universit\"{a}t zu Berlin, 
Invalidenstr. 110, D-10115 Berlin, Germany}
\end{center}

\begin{abstract} 

The dynamic sructure factor as the basic quantity describing the collective excitations in a fluid is considered. We consider the cases of neutral and Coulombic fluids. The classical method of moments is applied to construct the dynamic structure factor satisfying all known sum rules. An interpolational formula is found which expresses the dynamic characteristics of a classical or quantum fluid system in terms of its static correlation parameters. The analytical results based on the theory of moments are compared with Molecular dynamics data for various model systems.

\hspace*{0.5cm}

\end{abstract}

\section{Introduction}

In the past there has been considerable interest in the time dependence of correlation functions or equivalently of the frequency dependence of structure factors. These functions has been studied in neutral and Coulomb fluids both theoretically and by molecular-dynamic simulations \cite{Hansen1}. Under a neutral fluid we understand here a fluid of particles interacting via a short-ranged potential. That means the termin neutral fluids includes such nonneutral systems as dusty plasmas \cite{Thomas} and charged colloidal suspensions \cite{Farouki} where the interaction between the charged particles of one subsystem is screened by the motion of particles from another subsystem. 

Several approaches are devoted to the study of dynamic properties of strongly interacting fluid systems. In a rather incomplete list we mention the approaches in Refs. \cite{Hansen,Suttorp} based on the memory function formalism, the approaches based on the theory of moments \cite{Adamyan,Ortner}, and the approach based on the quasilocalized charge approximation \cite{Kalman}. It is interesting to note that all these approaches succeeded by exploiting the method of collective variables \cite{Ortner99} in  various modifications.

This paper gives a short overview of the application of the method of moments to the determination of dynamic properties of coupled fluid systems. As the main quantity describing the dynamics of a systems we consider the dynamic structure factor.
The dynamic structure factor may be measured in scattering experiments. The peaks in the dynamic structure factor determine the collective excitations of the system. There may propagate different collective excitations depending on the type of the system (neutral or Coulomb). Generally speaking in neutral fluids we deal with sound modes whereas the plasma mode is a finite frequency mode. The different behavior of the modes is connected with the different behavior of the interaction potential Fourier transform at small wavenumbers $k$. In a neutral fluid the Fourier transform is finite, in the Coulomb case it diverges for $k \to 0$. 


\section{Dynamic properties of neutral fluids}\label{neutral}

We consider a system of $N$ particles of one species with masses $m$ and interacting via a pair potential $V(r)$. The Fourier transform of the interaction potential  satisfies the inequality $V(k=0) < \infty$. The Hamiltonian of the neutral fluid reads:

\ba\label{Hamil}
H=\sum_{i=1}^{N} \frac{{p}_i^2}{2m}+\frac{1}{2}\sum_{i \neq j} V(\DK{x}_i-\DK{x}_j),
\ea
where $p_i$ is the $i$th particle momentum.
In what follows we will use a classical notation, though all the calculations are easily generalized to the quantum case.

Define the particle density and its Fourier transform
\ba\label{density}
n(\DK{r},t)=\sum_i \delta(\DK{r}-\DK{x}(t))\,,~~n_{\DK{k}}(t)=\sum_i e^{i \DK{k} \cdot \DK{x}_i(t)}
\ea
,the density-density correlation function
\ba\label{corr}
g(\DK{r},t)=\left \langle n(\DK{r},t)n(0,0) \right \rangle\,.
\ea
and the dynamic structure factor
\ba
S(\DK{k},\om)=\frac{1}{2\pi n} \int_{-\infty}^{\infty} e^{i(\om t-\DK{k} \cdot \DK{r})} g(\DK{r},t) dt \, d \DK{r}
\ea
In order to construct the dynamic structure factor as a central function for the determination of the dynamic properties of the system it is useful to consider the frequency moments of the dynamic structure factor:
\ba\label{moments}
M_n(k)=\int_{-\infty}^{\infty} \om^n S(\om,\DK{k})
= \frac{ i^n}{N} \left \langle \frac{d^n}{d t^n} n_{\DK{k}}(t) n_{-\DK{k}}(0) \right \rangle_{t=0}
\ea
Due to the parity of the structure factor all moments with odd numbers are equal to zero. The zeroth and second moments read
\ba\label{M0}
M_0(k)=S(k)\,, \\
M_2(k)=\frac{k^2}{m} k_BT\,.
\ea
where $S(k)=(1/N) \left \langle n_{\DK{k}} n_{-\DK{k}} \right \rangle$ is the  static structure factor of the fluid. 
The fourth moment includes particle correlations and reads,
\ba\label{M4}
M_4(k)&=&3k^4{(k_BT)^2}/{m^2}+M_4^{\rm pot}(k) \nonumber\\
M_4^{\rm pot}(k)&=&\frac{N}{Vm^2}k^4V(k)k_BT+\frac{1}{Vm^2}\\
\sum_{\DK{q}\neq -\DK{k}}&& \left[S(\DK{k}+\DK{q})-S(q)\right]\,(\DK{k}\cdot \DK{q})^2 k_B T\, V(q)\,.
\ea
The Nevanlinna formula of the classical theory of moments \cite{Adamyan} expresses the dynamic structure factor
\BE{nev}
S(\DK{k},z)= \frac{1}{\pi}\, {\rm Im}\;\frac{E_{n+1}(\DK{k},z)+q_n(\DK{k},z)E_n(\DK{k},z)}{D_{n+1}(\DK{k},z)+q_n(\DK{k},z)D_n(\DK{k},z)}
\EE
in terms of a function $q_n=q_n(\DK{k},z)$ analytic in the upper half-plane ${\rm Im}\,z>0$ and having a positive imaginary part there ${\rm Im}\,q_n(\DK{k},\om+i\eta)>0,\,\eta>0$, it also should satisfy the limiting condition: $\left( q_n(\DK{k},z)/z \right) \to 0$ as $z \to \infty$ within the sector $\theta < {\rm arg}(z)<\pi-\theta$.  The polynomials $D_n$ (and $E_n$) can be found in terms of the first $2n$ moments as a result of the Schmidt orthogonalization procedure. The first orthogonal polynomials read \cite{Adamyan}
\BEQ{poly}
D_1&=&z\,,~~D_2=z^2-\om_1^2\,,~~D_3=z(z^2-\om_2^2)\,,\\
E_1&=&M_0\,,~~E_2=M_0 z\,,~~~~~~E_3=M_0(z^2+\om_1^2-\om_2^2)\,,
\EEQ
where $\om_1^2(\DK{k})={M_2(\DK{k})}/{M_0(\DK{k})}$ and $\om_2^2(\DK{k})={M_4(\DK{k})}/{M_2(\DK{k})}$.
Consider first the approximation $n=1$ leading to the correct frequency moments $M_0$ and $M_2$. Using the Nevanlinna formula Eq. (\ref{nev}) we obtain ($q_1=q_{1,r}+i q_{1,i}$),
\BE{n=1}
S(\DK{k},\om)=\frac{S(k)}{\pi}  \, \, \frac{q_{1,i}(\DK{k},\om) \om_1^2}{\left[\om^2-\om_1^2(\DK{k})+q_{1,r}(\DK{k},\om)\om\right]^2+q_{1,i}^2(\DK{k},\om)\om^2}\,.
\EE
We have no phenomenological basis for the choice of that function $q_1(z)$ which would provide the exact expression for $S(\DK{k},z)$. We mention that the physical meaning of the function $h_1(z)=-i q_1(z)$ is that of a memory function  since from Eq. (\ref{n=1}) it follows that the inverse Fourier transform of the function $C(\DK{k},z)=(1/i\pi)\int_{-\infty}^{\infty}S(\DK{k},\om)/(z-\om)$ obeys the equation 
\BE{memory}
\frac{\partial^2 C(\DK{k},t)}{\partial t^2}+\om_1^2C(\DK{k},t)+\int_0^t ds\, h_1(\DK{k},t-s) \frac{\partial C(\DK{k},s)}{\partial s}=0 \,.
\EE
A simple approximation is to put the function $q_1(z)$ equal to its static value
$q_1(z)=q_1(0)=i \nu(\DK{k},)$
and Eq. (\ref{memory}) simplifies to the equation of a damped oscillator with frequency $\om_1$ and damping constant $\nu$.
\BE{damposc}
\frac{\partial^2 C(\DK{k},t)}{\partial t^2}+\om_1^2C(\DK{k},t)+\nu(\DK{k},) \frac{\partial C(\DK{k},t)}{\partial t}=0 \,.
\EE
From Eq. (\ref{damposc}) follows the dispersion relation of collective excitations in a classical neutral fluid, $\om_c^2(\DK{k})=\om_1^2(\DK{k})=\frac{M_2(\DK{k})}{M_0(\DK{k})}=\frac{k^2 k_B T}{m S(k)}$. The corresponding generalization to the quantum case ($T=0$) reads $\om_0(\DK{k})=\frac{\hbar k^2}{2 m S(k)}$ \cite{Feynman}. 

Consider now the long-wavelength behavior $k \to 0$. In this case the static structure factor $S(k \to 0)=n k_B T \kappa_T$ is determined by the compressibility $\kappa_T=-({1}/{V}) \left({\partial V}/{\partial P} \right)_T$. Then the dispersion relation reads
\ba
\om_c^2(k)=u^2k^2\,,
~~~u^2=\left(\frac{\partial P}{\partial \rho} \right)_T = \frac{v_s^2}{\gamma}\,,~~\gamma=\frac{c_p}{c_v}\,,
\ea
which differs from the familiar dispersion equation for the sound wave by the factor $\gamma$. For a model of independent oscillators: $c_p=c_v$ and $\gamma=1$. Therefore the above approximation for the static structure factor based on the Nevanlinna equation with $n=1$ represents the model of independent damped quasiparticles.

To go beyond this approximation one has to choose the 3-moment approximation $n=2$ in the Nevanlinna hierarchy reproducing the moments $M_0$, $M_2$ and $M_4$. Within this approximation and choosing $q_2(\DK{k},\om)=h(\DK{k})$ we obtain the following expression for the dynamic structure factor:
\BE{n=2}
S(\DK{k},\om)= \frac{S(k)}{\pi} \, \, \frac{h(\DK{k}) \om_1^2(k) \left(\om_2^2(k)-\om_1^2(k)\right)}{\om^2(\om^2-\om_2^2)^2+h^2(\DK{k})(\om^2-\om_1^2)^2}\,,
\EE
where $h(\DK{k})$ has to be taken from the relation 
\BE{h(k)}
h(k)={(\om_2^2-\om_1^2)}/{\nu(k)}=({S(k)}/\pi) ({(\om_2^2/\om_1^2-1)}/{S(k,0)})
\EE
in order to satisfy the exact low freqency behavior $S(\DK{k},0)$. The value $S(\DK{k},0)$ may be taken from elastic scattering experiments, from another theory or it may be used as a fit parameter.

Consider again the long wave-length limit $k \to 0$. Then the frequencies
\ba
\om_1^2(k)=u^2k^2\,,~~~u^2=\left(\frac{\partial P}{\partial \rho}\right)_T\\
\om_2^2(k)=v^2k^2\,,~~~v^2=\frac{n}{m}V(0)+3\frac{k_BT}{m}\,.
\ea
At small temperatures $k_BT \ll nV(0)$ we have $u^2=v^2$ and we obtain the dynamic structure factor for a ``classical'' fluid at low temperature
\ba
S(k,\om)=\frac{\pi k_B T}{m u^2} \left\{\delta(\om-ku)+\delta(\om+ku) \right\} \,,
\ea
representing undamped sound waves. The corresponding generalization to a quantum fluid reads
\ba
S(k,\om)=\frac{\pi \hbar k}{m u (1-\exp(-\hbar ku/k_BT))} \left\{\delta(\om-ku) +
 e^{-\hbar ku/k_BT} \delta(\om+ku) \right\},
\ea
At zero temperature the system may only absorb energy and we obtain the simple equation for the dynamic structure factor.
\ba
S(k,\om)=\frac{\pi \hbar k}{mu} \delta(\om-ku)\,.
\ea

\section{Dynamic properties of Coulomb  fluids}\label{Coulomb}

Consider a one component plasma (OCP) consisting of $N$ particles with charges $Ze$ and masses $m$ interacting via the Coulomb potential $V_c(r)={Z^2 e^2}/{r}$ and embedded in a neutralizing homogeneous background. The classical OCP may be characterized by the coupling parameter $\Gamma=\frac{e^2}{a k_BT}$, $a=(3V/4\pi N)^{1/3}$ being the Wigner-Seitz radius. The quantum plasma has an additional parameter - the degeneration parameter $\theta=\sqrt{2mk_BT}/\hbar^2(3 \pi^2 n)^{2/3}$. In what follows for the case of simplicity we concentrate on a classical plasma. For $\Gamma \ll 1$ we deal with an ideal (or Vlasov) plasma, for $\Gamma \gg 1$ the plasma is called a strongly coupled one. The Vlasov approximation takes into account only the mean field part of the interaction and the dispersion relation for the longitudinal plasmons is predicted as $\om_c^2(k)=\om_p^2 \left(1+3\frac{k^2}{k_D^2}\right)$ with the plasma frequency $\om_p^2=\frac{4 \pi Z^2 e^2 n}{m}$ and the squared inverse Debye-length $k_D^2=\frac{4 \pi Z^2 e^2 n}{k_BT}$. The Vlasov theory predicts a strong positive dispersion of the plasmons, i.e., ${d\om}/{dk}>0$. However, in a coupled plasma the potential energy plays an important role and the Vlasov approximation is not longer valid. To construct the dynamic structure factor for a coupled plasma consider the { frequency moments} of the dynamic structure factor $S(k,\om)$. The frequency moments formally coincides with that of a neutral plasma (Eqs. (\ref{M0})-(\ref{M4})). The only difference is that the interaction potential of the neutral fluid has to be replaced by that of the Coulomb system. The application of the Nevanlinna formula leads then to a corresponding hierarchy of approximations for the dynamic structure factor. If one is interested in the structure factor of a quantum system again Eqs. (\ref{nev}) hold, if one replaces $S(\DK{k},\om)$ on the left hand side of the Eqs. (\ref{nev})  by the loss function $R(\DK{k},\om=\left[(1-\exp(-\beta\hbar\om))/\beta\hbar\om\right]S(\DK{k},\om)$. However, in the quantum case additional contributions to the zeroth and fourth frequency moment occur \cite{Adamyan,Ortner}.

Consider the 3-moment approximation Eq. (\ref{n=2}). If one is interested in the investigation of the high-frequency collective excitation spectrum only it is sufficient to neglect the function $h(k)$ since the damping (described by the function $h$) is small in strongly coupled plasmas. If one puts $h(k)=0$ Eq. (\ref{n=2}) provides the expression of the dynamic structure factor for a strongly coupled plasma obtained within the QLC approach \cite{Kalman}, if the thermal contributions may be neglected with respect to the correlation contributions. Within the simple approximation $h(k)=0$ the dynamic structure factor has $\delta$ peaks at the frequencies $\om_c$ which in the classical case are determined by the equation
\ba
\om_c^2(k)=\frac{M_4}{M_2}=
\om_p^2 \left(1+3\frac{k^2}{k_D^2} 
+ \frac{1}{N}
\sum_{\DK{q}\neq -\DK{k}} \left[S(\DK{k}+\DK{q})-S(q)\right]\,\frac{(\DK{k}\cdot \DK{q})^2}{k^2q^2}\right).
\ea
For $k \to 0$ the dispersion relation simplifies and we get 
$$\om_c^2(k)=
\om_p^2 \left(1+3\frac{k^2}{k_D^2}+\frac{4}{45}\frac{E_c}{k_BTn\Gamma}k^2a^2\right)$$
with  $E_c$ being the correlation energy density. Using the simple estimation $\frac{E_c}{k_BTn}=-0.9 \Gamma$ valid in the strong coupling regime one obtains the dispersion relation $\om_c^2(k)=
\om_p^2 \left[1+k^2a^2(-0.08+\Gamma^{-1})\right]$ and one predicts a negative dispersion for $\Gamma>13$.

\begin {figure} [h]
\unitlength1mm 
  \begin{picture}(120,100)
\psfig{figure=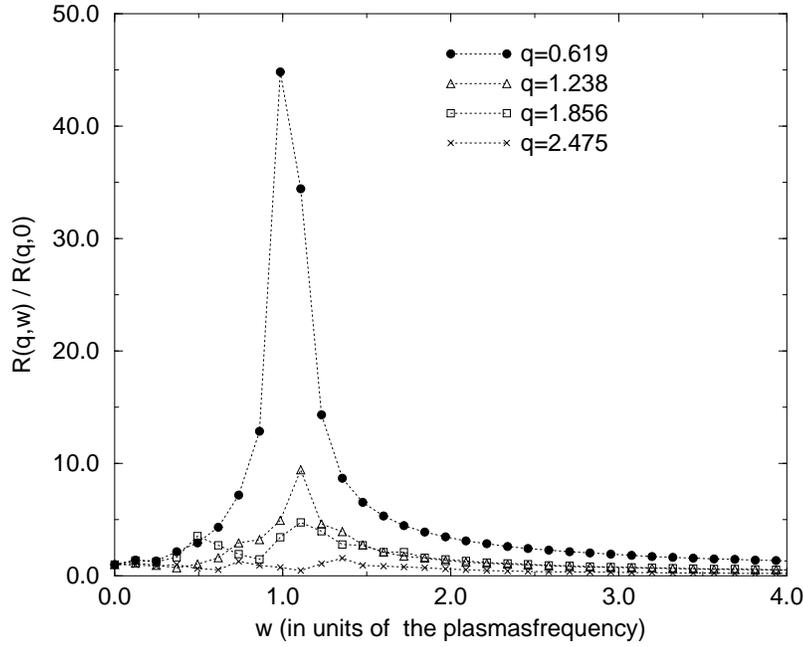,width=12.0cm,height=10.0cm,angle=0}
 \end{picture}\par 
\caption{\label{G10th1} {The simulation data for the loss function $R(q,\omega)$ versus frequency $\omega/\omega_p$  for different wavevectors $q=ka$ at $\Gamma=10$ and $\theta=1$.}}
\end{figure}

\begin {figure} [h]
\unitlength1mm 
  \begin{picture}(120,100)
\psfig{figure=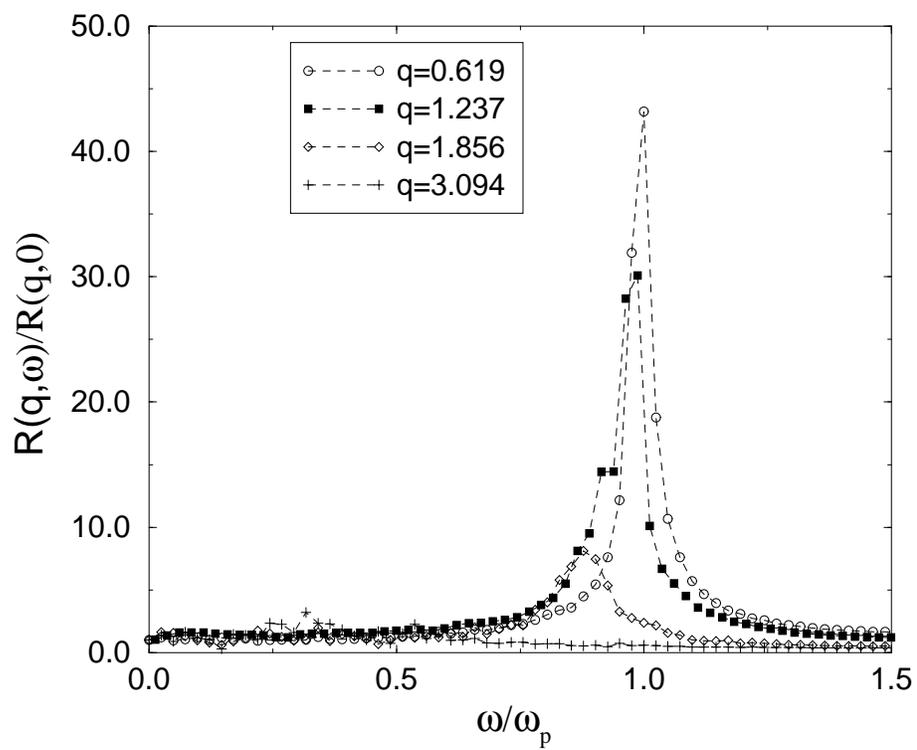,width=12.0cm,height=10.0cm,angle=-90}
 \end{picture}\par 
\caption{\label{G100th50} {Same as in Fig. 1 at $\Gamma=100$ and $\theta=50$}}
\end{figure}

To study the dynamic structure factor one has to go beyond the simple approximation $h=0$. To satisfy the low frequency behavior one may choose the approximation Eq. (\ref{h(k)}). To check the quality of the predictions from our approximation molecular dynamic simulations have been performed for comparison \cite{Ortner}. The semiclassical simulations were performed to model a quantum gas of 250 electrons moving in a cubic box with periodic boundary conditions.The thermal equilibrium was established by a Monte Carlo procedure. A detailed description of the semiclassical model used in the simulations may be found elsewhere \cite{Ortner}. In Figs. \ref{G10th1} and \ref{G100th50} we have plotted the loss function $ R(q,\omega) $ ($ q=ka $) for various values of wavenumbers $q$ for the cases of strong  ($ \Gamma=10$ ) and very strong coupling ($ \Gamma=100$ ) \cite{Ortner}. In both cases we obtain a sharp plasmon peak at small q values, with increasing wavenumber the plasmon peak widens. Almost no dispersion has been observed at $\Gamma=10$. This is in good agreement with the above estimation for the critical value $\Gamma=13$ separating regimes with positive dispersion from that with negative dispersion. For the case of very strong coupling $\Gamma=100$ we have found a strong negative dispersion.
In Figs. \ref{Nev1} and \ref{Nev2} we present the results of the MD data and compare them with our analytical approximation Eqs. (\ref{n=2}) and (\ref{h(k)}). To calculate the parameters  $\om_1(k)$ and $\om_2(k)$ we have used the static structure factor obtained from the HNC equations. The value $S(k,0)$ determining the parameter $h(k)$ might be taken from the MD simulations. However, the dynamic structure factor at the zero frequency can be obtained with the necessary accurazy only from long time simulations. Alternatively we have choosen the value $S(k,0)$ to fit the model to the MD data. It should be mentioned that the value $S(k,0)$ mainly determines the width of the plasmon peak, the peak position is quite insensitive to the choice of the value $S(k,0)$.  From the figures it can be seen that there is a reasonable agreement between the MD data and the present approxiamtion based on the sum rules. The peak position is reproduced with high accuracy, the agreement in the width of the peaks is less satisfactory. One concludes  that the static approximation $q_2(k,\om)=ih(k)$ undersetimates the damping of the quasiparticles.

\section{Conclusions}

In this paper we have shown that the application of the classical theory of moments gives a satisfactory description of many properties of neutral and Coulomb fluids. The Nevanlinna formula generates   approximate expressions for the dynamic structure factor in terms of their static correlations. The quality of the Nevanlinna expression mainly depends on the quality of the model used to calculate the static properties of the fluid. The presented results may be improved by a specification of the interpolation function $q_2(\DK{k},\om)$.

In conclusion, the present approach has been also used to calculate the dynamic structure factor of two-dimensional electron gas \cite{OT92}, of binary ionic mixtures \cite{AT91} and of two-component plasmas \cite{ATMV93}. It had been extended to magnetized plasmas \cite{ORT94} and can be generalized to calculate partial dynamic structure factors. Here, the matrix form of the Nevanlinna formula becomes helpful.

\newpage

\begin{center}
{\bf FIGURE CAPTIONS}
\end{center}

\begin {figure} [h]
\unitlength1mm 
  \begin{picture}(120,100)
\psfig{figure=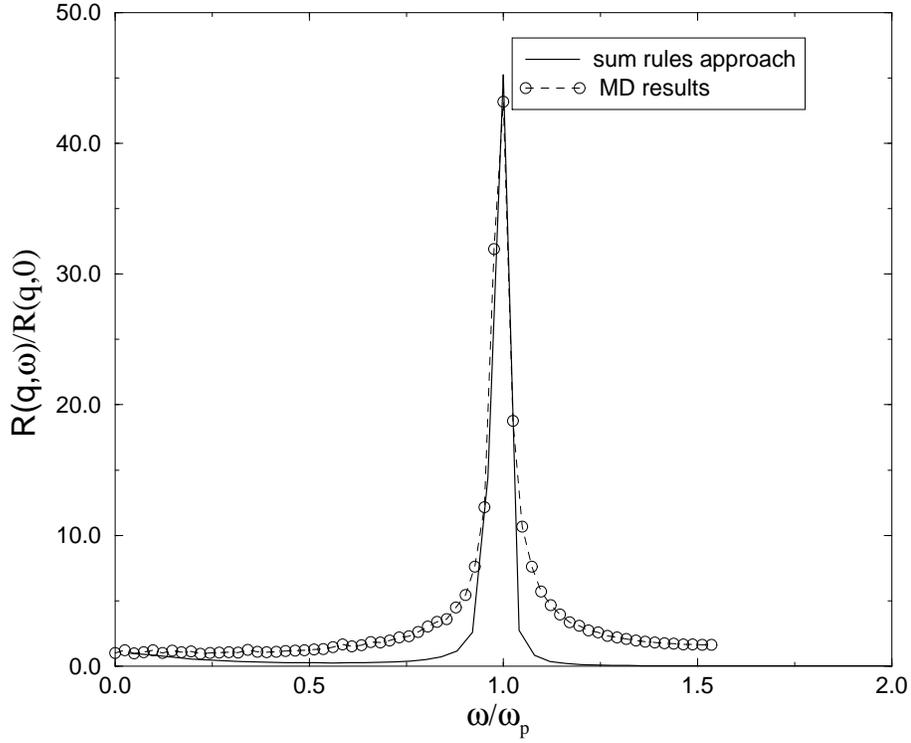,width=12.0cm,height=10.0cm,angle=-90}
 \end{picture}\par 
\caption{\label{Nev1} {Comparison of the loss function  $R(q,\omega)$ within the present sum rules approach (Eqs. (\ref{n=2} and (\ref{h(k)}) with $S(k,\om)$ replaced by $R(k,\om)$) versus frequency $\omega/\omega_p$ with the corresponding MD data at $\Gamma=100$ and $\theta=50$ for wavevector $q=0.619$.
.}}
\end{figure}

\begin {figure} [h] 
\unitlength1mm
  \begin{picture}(120,100)
\psfig{figure=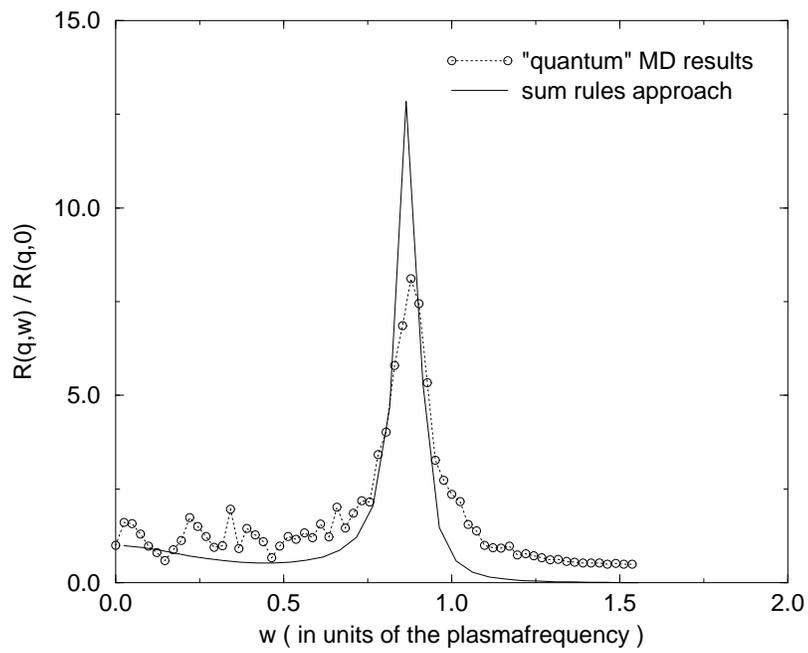,width=12.0cm,height=10.0cm,angle=0}
 \end{picture}\par 
\caption{\label{Nev2} { same as Fig.\ref{Nev1}; at $\Gamma=100$ and $\theta=50$ for wavevector $q=1.856$, 
{\cite{Ortner}}
.}}
\end{figure}

\begin{description}

\item[(Figure 1)] The simulation data for the loss function $R(q,\omega)$ versus frequency $\omega/\omega_p$  for different wavevectors $q=ka$ at $\Gamma=10$ and $\theta=1$.
\item[(Figure 2)] Same as in Fig. 1 at $\Gamma=100$ and $\theta=50$.
\item[{Figure 3}] Comparison of the loss function  $R(q,\omega)$ within the present sum rules approach (Eqs. (\ref{n=2} and (\ref{h(k)}) with $S(k,\om)$ replaced by $R(k,\om)$) versus frequency $\omega/\omega_p$ with the corresponding MD data at $\Gamma=100$ and $\theta=50$ for wavevector $q=0.619$. 
\item[{Figure 4}] Same as Fig.\ref{Nev1}; at $\Gamma=100$ and $\theta=50$ for wavevector $q=1.856$, 
{\cite{Ortner}}
.
\end{description}

\newpage

\newpage

\newpage

\end{document}